\documentclass[aps,showpacs,preprintnumbers,eqsecnum,amsmath,amssymb]{revtex4}
\usepackage{graphics,epsfig}
\usepackage{graphicx}
\usepackage{dcolumn}
\usepackage{bm}

\begin{document}

\title{ Energy of general 4-dimensional stationary axisymmetric spacetime
in the teleparallel geometry}

\author{Shanxian Xu}
\author{Jiliang Jing} \thanks{Corresponding author, Email: jljing@hunnu.edu.cn}
\affiliation{ Institute of Physics and  Department of Physics, \\
Hunan Normal University,\\ Changsha, Hunan 410081, P. R. China }

 \begin{abstract}

The field equation with the cosmological constant term is derived
and the energy of the general 4-dimensional stationary
axisymmetric spacetime is studied in the context of the
hamiltonian formulation of the teleparallel equivalent of general
relativity  (TEGR). We find that, by means of the integral form of
the constraints equations of the formalism naturally without any
restriction on the metric parameters, the energy for the
asymptotically flat/de Sitter/Anti-de Sitter stationary spacetimes
in the Boyer-Lindquist coordinate can be expressed as
$E=\frac{1}{8\pi }\int_S d\theta d\phi\left(sin\theta
\sqrt{g_{\theta\theta}}+\sqrt{g_{\phi\phi}}-(1/\sqrt{g_{rr}})(
\partial{\sqrt{g_{\theta\theta} g_{\phi\phi}}}/\partial
r)\right)$. It is surprised to learn that the energy expression is
relevant to the metric components $g_{rr}$, $g_{\theta\theta}$ and
$g_{\phi\phi}$ only. As examples, by using this formula we
calculate the energies of the Kerr-Newman  (KN), Kerr-Newman
Anti-de Sitter (KN-AdS), Kaluza-Klein, and Cveti\v{c}-Youm
spacetimes.

\vspace{0.2cm} \vspace*{0.4cm}

\end{abstract}


\maketitle

\section{Introduction}

With the advent of the general relativity (GR), the definition for
the gravitational energy continues to be one of the most active
areas of research in gravitational physics. It is well known that
the spacetime metric in general relativity describes both the
background spacetime structure and the dynamical aspects of the
gravitational field, but no natural way is known to decompose it
into its ``background'' and ``dynamical'' components
\cite{misner}. Thus, the notion of energy can not be obtained
without a corresponding decomposition of the spacetime metric.
Although a definition of the energy cannot be found locally, some
authors defined the pseudo-tensor of energy-momentum density
(\cite{ap}- \cite{ld}), and the quasilocal energy (\cite{brown}-
\cite{bose}) for the gravitational field. However, these
definitions have some defects. Say, if we consider the rotating
spacetime, the quasilocal energy can only be applied to the case
of the slow rotation spacetimes.

Recently, in the context of the teleparallel gravity (\cite{01}-
\cite{grqc0501017}), some authors re-examined the gravitational
energy problem and attempted to obtain the expression for the
gravitational energy in the form of a true spacetime and gauge
tensor (\cite{001755MPA04}- \cite{024021PRD62}). Teleparallel
gravity is not a new theory for the  gravitational field because
M\"{o}ller \cite{01} revived Einstein's GR by constructing a field
theory and reconsidered a Lagrangian formulation for absolute
parallism in 1962. Teleparallel gravity is just an alternative
geometrical formulation of GR, which is formulated on the
Weitzenb\"{o}ck spacetime \cite{Weitzenbock}. This gravitational
theory is characterized by the vanishing curvature tensor
(absolute parallism) and the nonvanishing torsion tensor. In this
theory, the dynamical field quantities are correspond to
orthogonal tetrad field $e_{a \mu}$
 ($a$ and $\mu$ are SO (3, 1) and spacetime indices,  respectively)
and the affine connection
${\stackrel{*}{\Gamma}}{}^{\lambda}{}_{\mu \nu}=e^{a\lambda}
\partial_\mu e_{a\nu}$.

Teleparallel gravity is commonly used to denote the general
three-parameter theory \cite{003524PRD19}. We consider only the
teleparallel equivalent of general relativity (TEGR), a theory
obtained for a specific choice of these parameters. In Refs.
(\cite{000335JMP35}- \cite{0504077}), the authors have established
a definition for the energy of the gravitational fields in the
framework of Schwinger's time gauge condition
 \cite{000800PR130} in terms of TEGR. Recently, in
Ref. \cite{grqc0205028}, by using this definition, Jos$\acute{e}$
and Karlucio have calculated the gravitational energy of Kerr and
Kerr Anti-de Sitter spacetime. However, we should note that the
Lagrangian density (Eq. (2.1) in Ref. (\cite{grqc0205028})) does
not include the cosmological constant term. Thus, it seems that
the definition of the energy educed from the Lagrangian density
can not be applied to the Kerr anti-de Sitter spacetime directly.

In this paper we first derive the field equation with the
cosmological constant. Then, we will generalize the definition of
the gravitational energy into the case with the cosmological
constant and the matter fields and obtain the general energy
expression for stationary axisymmetric spacetime. Then, as
examples,  we calculate the energy of the KN and KN-AdS
spacetimes, stationary Kaluza-Klein and rotating Cveti\v{c}-Youm
spacetimes enclosed by a arbitrary two-sphere. Furthermore, for
comparing with known results obtained by using other method, we
discuss the gravitational energy in some special cases, such as
$r\rightarrow\infty$, $r=r_{+}$, $a\rightarrow 0$ and slow
rotation approximation.

Notation: spacetime indices $\mu,  \nu,  . . . $ and SO (3, 1)
indices $a,  b,  . . . $ run from 0 to 3.  Time and space indices
are indicated according to $\mu=0, ~i, \;\;a= (0),~  (i)$. The
tetrad field $e^a\, _\mu$ yields the definition of the torsion
tensor: ${\stackrel{*}{T}}{}^a\, _{\mu \nu}=\partial_\mu e^a\,
_\nu-\partial_\nu e^a\, _\mu$.  The flat Minkowski spacetime
metric is fixed by $\eta_{ab}=e_{a\mu} e_{b\nu}g^{\mu\nu}=  (-,
~+, ~+, ~+)$.  The tangent space indices are raised and lowered
with the Lorentzian metric $\eta_{ab}$,  while the spacetime
indices are raised and lowered with the Riemannian metric
$g^{\mu\nu}$. All magnitudes related with GR will be denoted with
an over ``$\circ$", whereas magnitudes related with TEGR will be
denoted with an over ``$*$".

The organization of this paper is as follows. In Sec. II a general
energy expression for general stationary axisymmetric spacetime is
obtained. In Sections III, IV, V and VI the energy of the KN,
KN-AdS, Kaluza-Klein and Cveti\v{c}-Youm spacetimes are studied.
The last section is devoted to a summary.

\section{ General energy expression for general stationary axisymmetric  spacetime}

It is well known that, from the Riemann-Cartan type spacetime U4
\cite{000393RMP48} \cite{003524PRD19} follows two models of
space-time, one is the Riemann spacetime V4 of GR, another is the
Weitzenb\"{o}ck spacetime W4 of torsion gravity. The
Riemann-Cartan geometry is the basis of Poincar\'{e} gauge theory
 (PGT) \cite{grqc0501017}  \cite{000335JMP35} \cite{grqc0411119}. In this gauge fields,
the definitions of the curvature and torsion tensors are yielded
by the tetrad field $e_{a \mu}$ and the arbitrary spin affine
connection $\omega^{a}_{~b\mu}$,
\begin{eqnarray}\label{Rabcd}
R^a_{\ b \mu \nu }&=&\partial_\mu{\omega^{a}_{~b\nu}}
-\partial_\nu{\omega^{a}_{~b\mu}} +\omega^{a}_{~c\mu}
\omega^{c}_{~b\nu}
-\omega^{a}_{~c\nu} \omega^{c}_{~b\mu}, \\
T^a_{\ \mu \nu}&=&\partial_\mu{e^a_{\ \nu}}-\partial_\nu{e^a_{\
\mu}} +\omega^{a}_{~b\mu}e^b_{\ \nu}-\omega^{a}_{~b\nu} e^b_{\
\mu}.
\end{eqnarray}
The general spacetime connection $\Gamma^\lambda_{~ \mu \nu}$ is
defined by
\begin{eqnarray}
\Gamma^\lambda_{~\mu \nu}& =&e_a^{\ \lambda}\partial_\mu{e^a_{\
\nu}}+e_a^{\ \lambda}\omega^{a}_{~b\mu} e^b_{\ \nu}.
\end{eqnarray}
The spacetime and the tangent-space metrics are related by $g_{\mu
\nu}=e^a_{\ \mu} e^b_{\ \nu}\eta_{a b}$.
For the arbitrary spin
connection, there exists an identity
\begin{eqnarray}
\omega^{a}_{~b\rho}={\stackrel{\circ}{\omega}}{}^{a}_{~b\rho
}+{K}^{a}{}_{b\rho }=-\omega^{~a}_{b~\rho },
\end{eqnarray}
where  ${\stackrel{\circ}{\omega}}{}^{a}_{~b\rho }$ is the
Levi-Civita connection (the Ricci coefficient of rotation, the
spin connection of GR) which is metric compatible and torsion
free. It is related with the Christoffel connection
$\left\{^\lambda_{\mu\nu} \right\}$ of GR by
\begin{eqnarray}
{\stackrel{\circ}{\omega}}{}^{a}_{~b\mu }&=&e^a_{\ \lambda}\left\{
^\lambda_{\mu\nu}\right\} -e_b^{\
\lambda}\partial_\mu{e^a_{\ \lambda}},\\
\left\{ ^\lambda_{\mu\nu}\right\} &=&\frac{1}{2} g^{\lambda \rho}
(\partial{_\mu}g_{\nu \rho}+\partial{_\nu}g_{\mu \rho}-
\partial{_\rho}g_{\mu \nu}).
\end{eqnarray}
The contortion tensor $K_{ab\rho }$ is given by
\begin{eqnarray}
K_{ab\rho}&=&\frac{1}{2}e_a^{\ \mu}e_b^{\ \nu} K_{\mu \nu\rho }\nonumber\\
           &=&\frac{1}{2}e_a^{\ \mu} e_b^{\ \nu}
            (T_{\mu \nu \rho }-T_{\nu \mu \rho}-T_{\rho \mu \nu}),
\end{eqnarray}
with the torsion tensor
\begin{eqnarray}
T^\lambda_{\ \mu \nu}=\Gamma^{\lambda}_{\ \mu
\nu}-\Gamma^{\lambda}_{\ \nu \mu} =-T^\lambda_{\ \nu \mu}.
\end{eqnarray}

It is convenient to define a useful tensor $\Lambda^{a b c}$ by
\begin{eqnarray}
\Lambda^{a b c}=\frac{1}{4} (T^{a b c}+T^{b a c}-T^{c a b})
+\frac{1}{2} (\eta^{a c} T^b-\eta^{a b}T^c).
\end{eqnarray}
Then, we have
\begin{eqnarray}
\Lambda^{a b c} T_{a b c}=\frac{1}{4}T^{a b c}T_{a b
c}+\frac{1}{2}T^{a b c}T_{b a c}-T^a T_a,
\end{eqnarray}
where
\begin{eqnarray}
T_b=T^a_{\ \ a b}, \ \ \ \ \ \ \ \ T_{a b c}=e_b^{\ \mu}e_c^{\
\nu}T_{a \mu \nu}.\nonumber
\end{eqnarray}

Denoting the scalar curvature $R (e, \omega)=e^{a \mu}e^{b
\nu}R_{a b \mu \nu}$ and substituting the arbitrary spin
connection
 $\omega_{\rho a b}$
into the scalar curvature tensor, we get an identity
\cite{024021PRD62} \cite{000335JMP35}
\begin{eqnarray}
eR (e, \omega)=e{\stackrel{\circ}{R}}+e\Lambda^{a b c} T_{a b
c}-2\partial_\mu{ (eT^\mu)}.
\end{eqnarray}
This identity is the one of bases of the equivalence of TEGR with
Einstein's GR. The Einstein-Hibert Lagrangian without the
cosmological constant term is
\begin{eqnarray}
L_{GR}&=&ke{\stackrel{\circ}{R}}\nonumber\\
      &=&keR (e, \omega)-ke\Lambda^{a b c} T_{a b c},
\end{eqnarray}
where $2\partial_\mu{ (eT^\mu)}$ is a total derivative term and is
neglected, and $k=\frac{c^4}{16\pi G}\ $ ($c$ is light speed and
$G$ is the Newton's gravitational constant). Hereafter we will
take $G=c=1$.

In what following, we will study the case of the Lagrangian
density with the cosmological constant term. Then, the Lagrangian
density in the torsion is given by \cite{024021PRD62}
\cite{grqc0411119}
\begin{eqnarray} \label{Lagran}
L=e (L_\circ+aR^{a b \mu \nu}R_{a b \mu \nu}-L_M),
\end{eqnarray}
with
\begin{eqnarray}
L_\circ=k R (e, \omega)-k \Lambda^{a b c} T_{a b c}+2k\lambda,
\end{eqnarray}
where $\lambda$ is the cosmological constant, $aR^{a b \mu \nu}$
are Lagrange multipliers introduced to ensure the teleparallelism
condition in the variational formalism and $L_M$ is the Lagrangian
of matter fields. The lagrangian $L_\circ$ is then an alternate
``more geometric" interpretation of $L_M$.

Define the canonical stress-energy tensor and spin density in the
forms of \cite{grqc0411119}
\begin{eqnarray}
T^a_{\ \mu}=\frac{1}{2e}\frac{\delta {\mathcal{L}}_M}{\delta
e_a^{\ \mu}},\ \ \ \ \ \ \ \ \ \ \ S^{\ \
\rho}_{ab}=\frac{1}{e}\frac{\delta {\mathcal{L}}_M}{\delta
\omega_{\ \ \rho}^{ab}},\nonumber
\end{eqnarray}
where ${\mathcal{L}}_M=e L_M$. Then, the field equation can be
expressed as
\begin{eqnarray}
&&{\stackrel{\circ}{G}}{}_{\mu\nu}+\lambda g_{\mu\nu}
=\tau_{\mu\nu}+T_{\mu\nu},\label{fangcheng}\\
&&D_\lambda R^{ab\rho\lambda}=S^{ab\rho},
\end{eqnarray}
where $\tau_{\mu\nu}=-2a (R^{ab}_{\ \ \rho\mu}R_{ab\ \nu}^{\ \
\rho} -\frac{1}{4}g_{\mu\nu}R^{ab}_{\ \ \rho\lambda}R_{ab}^{\ \
\rho\lambda})$ and $D_\lambda$ is the covariant derivative that
acts with $\omega^{\ \ \rho}_{ab}$ on the tangent space indices
and with $\{^\lambda_{\mu\nu}\}$ on the spacetime indices. We can
refer to other forms of the field equation
 \cite{000393RMP48} \cite{003524PRD19} \cite{044010PRD64} \cite{000335JMP35}.

In TEGR,  the fundamental field variable describing gravity is a
quadruple of parallel tetrad field $e_{a \mu}$ which is on the
Weitzenb\"{o}ck spacetime. In this geometrical structure, define
the so called the Weitzenb\"{o}ck connection
\begin{eqnarray}
{\stackrel{*}{\Gamma}}{}^{\lambda}{}_{\mu \nu}&=&e_a^{\
\lambda}\partial{_\mu}e^a_{\ \nu},
\end{eqnarray}
as a consequence of the absolute parallelism condition (i. e., the
Weitzenb\"{o}ck covariant derivative  of the tetrad field vanishes
identically: ${\stackrel{*}{\bigtriangledown}}{}_\nu e^a_{\
\mu}\equiv
\partial_\nu e^a_{\ \mu}-{\stackrel{*}{\Gamma}}{}^{\lambda}{}_{\mu \nu}e^a_{\ \lambda}=0$
). Then, we know that Weitzenb\"{o}ck spin connection vanishes
${\stackrel{*}{\omega}}{}^{a}_{~b\rho }=0$, the curvature of the
Weitzenb\"{o}ck connection also vanishes ${\stackrel{*}{R}}{}^a_{\
b \mu \nu }=\partial_\mu{{\stackrel{*}{\omega}}{}^{a}_{~b\nu}}
-\partial_\nu{{\stackrel{*}{\omega}}{}^{a}_{~b\mu}}
+{\stackrel{*}{\omega}}{}^{a}_{~c\mu}
{\stackrel{*}{\omega}}{}^{c}_{~b\nu}
-{\stackrel{*}{\omega}}{}^{a}_{~c\nu}
{\stackrel{*}{\omega}}{}^{c}_{~b\mu}=0$, and the torsion of TEGR
reduce to
\begin{eqnarray}
{\stackrel{*}{T}}{}^{a}{}_{\mu \nu}\equiv
\partial_\mu{e^a_{\ \nu}}-\partial_\nu{e^a_{\ \mu}}.
\end{eqnarray}

In the absence of spinning sources (i.e., $S^{ab\rho}=0$), the
classical matter couples to the metric alone. Thus, in the
framework of the teleparallel geometry in PGT, we get
${\stackrel{*}{R}}{}_{ab}^{\ \ \rho\lambda}=0$ (namely, fix the
Poincar\'{e} gauge in terms of ${\stackrel{*}{\omega}}{}_{\ \
\rho}^{ab}=0$) as ground state solution and can obtain
$\tau_{\mu\nu}=0$, and above field equation (\ref{fangcheng}) can
reduce to the Einstein equation of GR, $L_\circ$ reduce to
$L_\circ=-k({\stackrel{*}{\Lambda}}{}^{abc}{\stackrel{*}{T}}{}_{abc}
-2\lambda)$, and L simplifies to $L
(e_{a\mu})=-ke({\stackrel{*}{\Lambda}}{}^{abc}{\stackrel{*}{T}}{}_{abc}-2\lambda)-eL_M$.
Namely, the equivalence with general relativity holds certainly.
However, we should note that the gravitational couplings to
spinning matter field in the Riemann spacetime (in GR) and in the
Weitzenb\"{o}ck spacetime (in TEGR) are in general different.

The Lagrangian density (\ref{Lagran}) can be expressed as
\begin{eqnarray}
L(e_{a\mu})=-4ke{\stackrel{*}{\Lambda}}{}^{a0k}{\stackrel{\cdot}{e}}{}_{ak}
+4ke{\stackrel{*}{\Lambda}}{}^{a0k}\partial_ke_{a0} -k
e{\stackrel{*}{\Lambda}}{}^{aij}T_{aij}+2\lambda k e - eL_M,
\end{eqnarray}
where the dot indicates a time derivative, and
${\stackrel{*}{\Lambda}}{}^{a0k}={\stackrel{*}{\Lambda}}{}^{abc}e_b^{\
0}e_c^{\ k}, \
{\stackrel{*}{\Lambda}}{}^{aij}={\stackrel{*}{\Lambda}}{}^{abc}e_b^{\
i}e_c^{\ j}$. The momentum canonically conjugated to $e_{a \mu}$
is given by
 \cite{084014PRD64} \cite{124001PRD65}
\begin{eqnarray}
\Pi^{a k}=-4 k e {\stackrel{*}{\Lambda}}{}^{a 0 k},
\end{eqnarray}
and the full expression  of $\Pi^{a k}$ can be expressed as
\begin{eqnarray}
\Pi^{a k}&=&k e[g^{0 0} (-g^{k j}{\stackrel{*}{T}}{}^a_{\ 0
j}-e^{a j}{\stackrel{*}{T}}{}^k_{\ 0 j}+2e^{a
k}{\stackrel{*}{T}}{}^j_{\ 0 j}) +g^{0 k} (g^{0
j}{\stackrel{*}{T}}{}^a_{\ 0 j}+e^{a j}{\stackrel{*}{T}}{}^0_{\ 0
j})\nonumber
\\&+&e^{a 0} (g^{0 j}{\stackrel{*}{T}}{}^k_{\ 0 j}+g^{k i}{\stackrel{*}{T}}{}^0_{\ 0 j}
-2 (e^{a 0}g^{0 k}{\stackrel{*}{T}}{}^j_{\ 0 j}+e^{a k}g^{0
j}{\stackrel{*}{T}}{}^0_{\ 0 j})-g^{0 i}g^{k
j}{\stackrel{*}{T}}{}^a_{\ i j}\nonumber
\\&+&e^{a i} (g^{0 j}{\stackrel{*}{T}}{}^k_{\ i j}-g^{k j}{\stackrel{*}{T}}{}^0_{\ i j})-2 (g^{0 i}
e^{a k}-g^{i k}e^{a 0}) {\stackrel{*}{T}}{}^j_{\ j i}].
\end{eqnarray}
In terms of the quantities on the spacelike hypersurface in the time
gauge condition, we have
\begin{eqnarray} \label{expession1}
\Pi^{ (0)k}&=&2keg^{i k}g^{j m}e^{ (l)}_{\ \ \ m} {\stackrel{*}{T}}{}_{ (l)ij}\nonumber\\
          &=&-2ke{\stackrel{*}{T}}{}^k.
\end{eqnarray}
Therefore, the energy is given by \cite{084014PRD64}
\cite{124001PRD65}
\begin{eqnarray} \label{EEE}
 E&=&P^{ (0)}=-\int_V{d^3x\partial_i{\Pi^{ (0) i}}}\nonumber\\
    &=&\frac{1}{8\pi }\int_V{d^3x\partial_j{ (e{\stackrel{*}{T}}{}^j)}}. \label{eg}
\end{eqnarray}

In the macrophysical limit  \cite{000393RMP48}, the dynamical
characterization of a continuous distribution of macroscopic
matter can be successfully achieved by energy-momentum (``momentum
current'') alone because spin averages out in this case (Indeed
the spin and the gradients of spin may cancel out when such an
average is performed). Thus, when we want to apply the field
equations in the macro-scopic domain, TEGR and GR is equivalent.
The energy-momentum tensor of matter is the source of the metric
tensor $g_{\mu \nu}$ of a Riemannian spacetime. Then, we can pick
out a set of tetrad $e_{a\mu}$ corresponding to spacetime metric
$g_{\mu\nu}$ and understand  (\ref{eg}) as the total energy
\cite{grqc0412055} \cite{0504077}.

In the following we will try to apply the expression  (\ref{EEE})
to general 4-dimensional stationary axisymmetric   spacetime and
to find a concisely formula of the energy. In Boyer-Lindquist
coordinates the general stationary axisymmetric  spacetime can be
written as  \cite{JingE}
\begin{eqnarray}\label{SP}
ds^{2}&=&g_{tt}dt^{2}+2g_{t\phi}dt d\phi+g_{rr}dr^{2}+g_{\theta
\theta}d\theta^{2} +g_{\phi \phi}d\phi^{2},
\end{eqnarray}
where $g_{\mu\nu}$ are functions of the coordinates $r$ and
$\theta$ only. It is well known, for a given spacetime metric
tensor $g_{\mu \nu}$, that there exists an infinite set of tetrad
fields that yield $g_{\mu \nu}$. Considering the Schwinger's time
gauge condition
 \cite{124001PRD65} \cite{grqc0205028} \cite{004242JMP36} \cite{006302JMP37},
a set of tetrads associated to the spacetime  (\ref{SP}) can be
expressed as
\begin{eqnarray} \label{matric-e}
e_{a\mu}= \begin{pmatrix}-A&0&0&0\cr
B\sin\theta\sin\phi&C\sin\theta\cos\phi & Dr\cos\theta\cos\phi &
-Fr\sin\theta\sin\phi\cr -B\sin\theta\cos\phi& C\sin\theta
\sin\phi & Dr\cos\theta\sin\phi &  Fr\sin\theta\cos\phi\cr
0&C\cos\theta & -Dr\sin\theta & 0\cr
\end{pmatrix},
\end{eqnarray}
where
\begin{eqnarray}  \label{coff}
A&=&\sqrt{-\left (g_{tt}-\frac{g_{t\phi}^2}{g_{\phi\phi}}\right)},
~B=-\frac{g_{t\phi}}{\sqrt{g_{\phi\phi}}sin\theta},
~C=\sqrt{g_{rr}}, ~ D=\frac{\sqrt{g_{\theta\theta}}}{r}, ~
F=\frac{\sqrt{g_{\phi\phi}}}{r sin\theta}.
\end{eqnarray}
The components of the torsion tensor obtained out of the tetrads
(\ref{matric-e}) are presented in the appendix. Thus, for an
arbitrary spacelike two-spheres, we know from Eq.  (\ref{EEE})
that the general energy expression of the asymptotically flat/de
Sitter/Anti-de Sitter stationary spacetimes (\ref{SP}) is given by
\begin{eqnarray}\label{def}
E=\frac{1}{8\pi }\int_S d\theta d\phi\left(sin\theta
\sqrt{g_{\theta\theta}}+\sqrt{g_{\phi\phi}}-\frac{1}{\sqrt{g_{rr}}}
\frac{\partial{\sqrt{g_{\theta\theta} g_{\phi\phi}}}}{\partial
r}\right).
\end{eqnarray}
It is interesting to note that the energy expression is very
simple and is only relevant to the metric components $g_{rr}$,
$g_{\theta\theta}$ and $g_{\phi\phi}$.

As examples, we now study the energies of the well-known
spacetimes, such as the KN, KN-AdS, Kaluza-Klein and
Cveti\v{c}-Youm spacetimes by using the formula  (\ref{def}).

\section{Energy of the KN spacetime}

The line element of the KN spacetime can be expressed as
\begin{eqnarray}
\label{1a}ds^{2}&=&-\frac{\Delta-a^{2}\sin^{2}\theta}
{\rho^{2}}dt^{2}+\frac{a (Q^2-2mr)\sin^{2}\theta}{\rho^{2}}dtd\phi\nonumber\\
&&+
\frac{\rho^{2}}{\Delta}dr^{2}+\rho^{2}d\theta^{2}+\frac{\Sigma^{2}\sin^{2}\theta}{\rho^{2}}d\phi^{2},\end{eqnarray}
with
\begin{eqnarray}
 \Delta=r^{2}+a^{2}+Q^{2}-2mr,~~
 \Sigma=\sqrt{ (r^2+a^2)^2-\Delta a^{2}\sin^{2}\theta},\nonumber
\end{eqnarray}
where $m$, $Q$ and $a$ represent the mass, electric and rotation
parameter of the spacetime , respectively. By using the Eq
 (\ref{def}) we find that the energy of the KN spacetime is given
by
\begin{eqnarray}\label{Ekn}
E_{KN} &&=\frac{1}{4}\left (\sqrt{r^2+a^2}
+\frac{r^2}{2a}\ln{\frac{
\sqrt{r^2+a^2}+a}{\sqrt{r^2+a^2}-a}}\right)\nonumber\\
&&+\frac{1}{4}\left (\frac{2 \sqrt{\alpha}}{i
a}-\frac{\sqrt{\alpha}
\partial_r{\Delta}}{i a \sqrt{\Delta}}\right)
E\left (\frac{i a}{r}, \frac{r \sqrt{\Delta}}{\sqrt{\alpha}}\right)\nonumber\\
&&+\frac{1}{4}\left (\frac{\sqrt{\alpha} \partial_r{\Delta}}{i a
\sqrt{\Delta}} -\frac{\sqrt{\Delta} \partial_r{\alpha}}{i a
\sqrt{\alpha}}\right) F\left (\frac{i a}{r},
\frac{r\sqrt{\Delta}}{\sqrt{\alpha}}\right),
\end{eqnarray}
where  $\alpha= (r^2+a^2)^2-\Delta a^2$ and the elliptic functions
$E (x, y)$ and $F (x, y)$ are defined by
\begin{eqnarray}
E (x, y)&=&\int^x_0{du \frac{\sqrt{1-y^2 u^2}}{\sqrt{1-u^2}}},\nonumber\\
F (x, y)&=&\int^x_0{du \frac{1}{\sqrt{1-u^2} \sqrt{1-y^2 u^2}}}.
\nonumber
\end{eqnarray}

Now, let us consider some special cases:

 (1)  In the asymptotic limit $r\rightarrow\infty$, from
Eq. (\ref{Ekn}) we obtain
\begin{eqnarray}
 E_{KN}=m \nonumber.
\end{eqnarray}
This is just the Amowitt-Deser-Misner (ADM) energy of the KN
spacetime.

 (2) The energy of the static  ($a=0$) Reissner-Nordstr\"{o}m
spacetime can be easily found  using Eq. (\ref{Ekn})
\begin{eqnarray}
 E_{RN}=
r \left(1-\sqrt{1-\frac{2m}{r}+\frac{Q^2}{r^2}}\right). \nonumber
\end{eqnarray}
When $Q=0$,  the energy $R_{RN}$ reduces to $E_{SH}=r
\left(1-\sqrt{1-\frac{2m}{r}}\right)$, which is the energy within
an arbitrary spacelike surface of fixed radius r in the
Schwarzschild spacetime  \cite{004242JMP36}.

 (3) If the surface is located at the event horizon of the KN spacetime,
 noting $ \Delta=0$ we find
\begin{eqnarray}\label{KKKr}
E_{KN} (r_+) &=&\frac{M_{irr}}{2}+\frac{12M_{irr}^2-a^2}{8 a} ln
\frac{2M_{irr}+a}{2M_{irr}-a},
\end{eqnarray}
where $r=r_+=m+\sqrt{m^2-a^2-Q^2}$ and
$M_{irr}=\sqrt{r_+^2+a^2}/2$.  If we take $Q=0$ Eq. (\ref{KKKr})
reduces to the result of the Kerr spacetime obtained in Ref.
\cite{124001PRD65}.

 (4) If we  consider the slow rotation approximation  (namely,
$\frac{a}{r}<<1$), we get
\begin{eqnarray}
 E_1&=&\frac{1}{4}\int_0^\pi{ (\rho
 +\frac{\Sigma}{\rho})\sin{\theta}}d\theta
   =r+\frac{a^2}{6r} (2+\frac{2m}{r}-\frac{Q^2}{r^2})+
   O (\frac{a^4}{r^4}),\nonumber\\
 E_2&=&\frac{1}{4}\int_0^\pi{-\frac{\sqrt{\Delta}\partial_r{\Sigma}}
 {\rho}
 \sin{\theta}}d\theta \nonumber\\
   &=&-\sqrt{r^2-2Mr+a^2+Q^2}+\frac{a^2}{6r} \sqrt{1-\frac{2m}{r}
   +\frac{a^2+Q^2}{r^2}}  (1+\frac{2m}{r}-\frac{2Q^2}{r^2})
 +O (\frac{a^4}{r^4}). \nonumber
\end{eqnarray}
Then, the energy of the slow rotating KN spacetime is shown by
\begin{eqnarray}
E_{KN}&=&E_1+E_2=r\left (1-\sqrt{1-\frac{2m}{r}+\frac{a^2+Q^2}{r^2}}\right )\nonumber\\
   &+&\frac{a^2}{6r}\left [2+\frac{2m}{r}-\frac{Q^2}{r^2}
   + (1+\frac{2m}{r}-\frac{2Q^2}{r^2})\sqrt{1-\frac{2m}{r}+\frac{a^2+Q^2}{r^2}}\right ]
   +O (\frac{a^4}{r^4}), \label{EKNN}
\end{eqnarray}
which is just the Brown-York quasilocal energy of the KN spacetime
 \cite{104027PRD60}. It is interesting to note that the term $-E_1$
corresponds exactly to $\varepsilon_0$ in Ref. \cite{104027PRD60}
which is a reference term proposed in Ref.  \cite{Hawking} for the
normalization of the energy with respect to a reference spacetime,
and the term $E_2$ yields $\varepsilon$ in Ref. \cite{104027PRD60}
.

\section{ The energy of  KN-AdS spacetime }

In Boyer-Lindquist coordinates the KN-AdS spacetime can be written
as  \cite{280CMP10} \cite{AP98-127}
\begin{eqnarray}
\label{2}ds^{2}&=&-\frac{\Delta-\Lambda_\theta
a^{2}\sin^{2}\theta}
{\rho^{2}}dt^{2}-\frac{2a[\Lambda_\theta (r^2+a^2)-\Delta]\sin^{2}\theta}{\chi\rho^{2}}dtd\phi\nonumber\\
&&+\frac{\rho^{2}}{\Delta}dr^{2}+\frac{\rho^{2}}{\Lambda_\theta}d\theta^{2}
+\frac{\Sigma^{2}\sin^{2}\theta}{\chi^2 \rho^{2}}d\phi^{2},
\end{eqnarray}
with
\begin{eqnarray}
 \Lambda_r&=&1+\frac{r^2}{l^2},\ \ \ \Lambda_\theta=1-\frac{a^2}{l^2}\cos^2 \theta,\ \ \ \ \chi=1-\frac{a^2}{l^2},\nonumber\\
 \Delta&=&\Lambda_r (r^{2}+a^{2})+Q^{2}-2mr, \ \ \ \ \ \ \ \ \ \ \ \rho=\sqrt{r^{2}+a^{2}\cos^{2}\theta},\nonumber\\
 \Sigma&=&\sqrt{\Lambda_\theta (r^2+a^2)^2-\Delta a^{2}\sin^{2}\theta} ,\nonumber
\end{eqnarray}
where $m$, $Q$ and $a$ are respectively mass, electric and
rotation parameters, and $l$ is the AdS radius which relates to
the cosmological constant by $\lambda=-3/l^2$.

By using the Eq  (\ref{def}), we find that the energy of the
KN-AdS spacetime can be described by
\begin{eqnarray}\label{AdS}
E_{AdS}
 &=&\frac{lr}{2a}E\left (\frac{a}{l},\frac{il}{r}\right)
 +\frac{\sqrt{\alpha}}{2ia \chi} E\left (\frac{ia}{r},
 \frac{r\sqrt{\beta}}{\sqrt{\alpha}}\right)
 +\frac{il}{4a\chi\sqrt{l^2-a^2}}
 \frac{\partial_r{\alpha}}{r}
 F\left (\frac{ia \sqrt{\Delta}}{\sqrt{\alpha}},
 \frac{\sqrt{\alpha (l^2+r^2)}}{r\sqrt{\Delta (l^2-a^2)}}\right)\nonumber\\
&+& \frac{il (\chi r-m)}{2a\chi \sqrt{l^2-a^2}}
\left\{r\left[F\left (\frac{r\sqrt{\chi
(r^2+a^2)}}{\sqrt{\alpha}},\frac{\sqrt{\alpha (r^2+l^2)}}
{r\sqrt{\Delta (l^2-a^2)}}\right)-F\left (1,\frac{\sqrt{\alpha
(r^2+l^2)}}
{r\sqrt{\Delta (l^2-a^2)}}\right)\right] \right.\nonumber\\
&+&\left.\frac{ (\alpha-r^2\beta)}{r\beta}\left[ \Pi\left
(\frac{\alpha}{r^2\beta}| \frac{r\sqrt{\chi
(l^2+a^2)}}{\sqrt{\alpha}}, \frac{\sqrt{\alpha (r^2+l^2)}}
{r\sqrt{\Delta (l^2-a^2)}}\right) -\Pi\left
(\frac{\alpha}{r^2\beta}| 1, \frac{\sqrt{\alpha
(r^2+l^2)}}{r\sqrt{\Delta (l^2-a^2)}}\right)\right]\right\},
\nonumber \\
\end{eqnarray}
where $\beta=\Delta-\frac{ (r^2+a^2)^2}{l^2},$ and the elliptic
function $\Pi (n |x,y)$ is defined by
\begin{eqnarray}
\Pi (n |x,y)&=&\int^x_0{du \frac{1}{ (1-n u^2)\sqrt{1-u^2}
\sqrt{1-y^2 u^2}}}. \nonumber
\end{eqnarray}

In the limit $a\rightarrow0$, we can find that Eq.  (\ref{AdS})
leads to the energy of the Reissner-Norstr\"om AdS spacetime,
$E_{RNAdS}= r (1-\sqrt{1-\frac{2m}{r}+\frac{Q^2}{r^2}+
\frac{r^2}{l^2}}\ )$.

 If we  consider the slow rotation approximation, we get
\begin{eqnarray}
 E_1&=&\frac{1}{4}\int_0^\pi{ (\frac{\rho}{\sqrt{\Lambda_\theta}}
 +\frac{\Sigma}{\chi\rho})\sin{\theta}}d\theta \nonumber\\
   &=&r (1+\frac{a^2}{3l^2})+\frac{a^2}{6r} (2+\frac{2m}{r}-\frac{Q^2}{r^2})+
   O (\frac{a^4}{r^4}),\nonumber\\
 E_2&=&\frac{1}{4}\int_0^\pi{-\frac{\sqrt{\Delta}\partial_r{\Sigma}}{\chi\rho\sqrt{\Lambda_\theta}}
 \sin{\theta}}d\theta \nonumber\\
   &=&-\sqrt{\Delta}+\frac{a^2}{6r} \sqrt{1-\frac{2m}{r}
   +\frac{a^2+Q^2}{r^2}+\frac{r^2+a^2}{l^2}}  (1+\frac{2m}{r}-\frac{2Q^2}{r^2})
   -\frac{2}{3}\frac{a^2}{l^2} \sqrt{\Delta}+O (\frac{a^4}{r^4}). \nonumber
\end{eqnarray}
Then, the energy of the slow rotating KN-AdS spacetime is shown by
\begin{eqnarray}
 E_{AdS}&=&E_1+E_2 \nonumber\\
   &=&r-r\sqrt{1-\frac{2m}{r}+\frac{a^2+Q^2}{r^2}+\frac{r^2+a^2}{l^2}}
   +\frac{r a^2}{3l^2}\left  (1-2\sqrt{1-\frac{2m}{r}+\frac{a^2+Q^2}{r^2}
   +\frac{r^2+a^2}{l^2}}\right )\nonumber\\
   &+&\frac{a^2}{6r}\left [2+\frac{2m}{r}-\frac{Q^2}{r^2}
   +\left (1+\frac{2m}{r}-\frac{2Q^2}{r^2}\right )\sqrt{1-\frac{2m}{r}+\frac{a^2+Q^2}{r^2}+\frac{r^2+a^2}{l^2}}\right ]
   ,\label{EEg}
 \end{eqnarray}
where the terms of order $O (\frac{a^4}{r^4})$ and higher are
neglected.

\section{ The energy of stationary Kaluza-Klein spacetime }
The Kaluza-Klein spacetime  is described by
 \cite{JingE} \cite{VFAZ}
\begin{eqnarray}
\label{1}ds^{2}&=&-\frac{1-Z}
{N}dt^{2}-\frac{2aZ\sin^{2}\theta}{N\sqrt{1-v^2}}dtd\phi\nonumber\\
&&+\frac{N\rho^{2}}{\Delta}dr^{2} +N\rho^{2}d\theta^{2}+\left[N
(r^2+a^2)+\frac{Z}{N}a^2\sin^{2}
{\theta}\right]\sin^{2}{\theta}d\phi^{2},\nonumber
\end{eqnarray}
where
\begin{eqnarray}
\rho&=&\sqrt{r^2+a^2\cos^{2}\theta},\ \ \ \ \ \ \ \ \ \Delta=r^2+a^2-2mr,\nonumber\\
Z&=&\frac{2mr}{\rho^{2}}, \ \ \ \ \ \ \ \ \ \ \ \ \ \ \ \ \ \ \ \
\ \ N=\sqrt{1+\frac{v^2Z}{1-v^2}},\nonumber
\end{eqnarray}
and $a$ and $v$ are the rotation parameter and the velocity of the
boost. The horizons, ADM mass, charge $Q$ and angular momentum $J$
of the spacetime  are given by
\begin{eqnarray}
r_\pm&=&m\pm\sqrt{m^2-a^2},\ \ \ \ \
M=m\left [1+\frac{v^2}{2 (1-v^2)}\right],\nonumber\\
Q&=&\frac{m v}{1-v^2},\ \ \ \ \ \ \ \ \ \ \ \ \ \ \ \
J=\frac{ma}{\sqrt{1-v^2}}.\nonumber
\end{eqnarray}
Considering
\begin{eqnarray}
\gamma&=&\frac{v^2}{1-v^2},\ \ \ \ \ \ \ \ \ \ \ \
\sigma=\sqrt{r^2+2\gamma
mr+a^2\cos^{2}\theta},\nonumber\\
\alpha&=& (r^2+a^2) (r^2+2\gamma mr+a^2)-\Delta a^2
= (r^2+a^2) (r^2+2\gamma mr)+2mra^2,\nonumber\\
\Sigma&=&\sqrt{ (r^2+a^2) (r^2+2\gamma mr+a^2)-\Delta
a^2\sin^{2}{\theta}}=\sqrt{\Delta
a^2\cos^{2}\theta+\alpha},\nonumber
\end{eqnarray}
we obtain
\begin{eqnarray}
g_{rr}=\frac{\rho\sigma}{\Delta},\ \ \ \ \ \ \ \
g_{\theta\theta}=\rho\sigma,\ \ \ \ \ \ \ \
g_{\phi\phi}=\frac{\Sigma^2 \sin^{2}{\theta}}{\rho\sigma}.
\end{eqnarray}
By using the Eq  (\ref{def}) we find that the energy of the
stationary Kaluza-Klein spacetime can be expressed as
\begin{eqnarray}
E_{KK}=\frac{1}{4}\int_0^\pi{\left (\sqrt{\rho}\sqrt{\sigma}+
\frac{\Sigma}{\sqrt{\rho}\sqrt{\sigma}}
 -\frac{\sqrt{\Delta}\ \partial_r{\Sigma}}{\sqrt{\rho}\sqrt{\sigma}}\right)
 \sin{\theta}}d\theta.
\end{eqnarray}
Unfortunately, the integral cannot be expressed in simple
functions. But we can discuss some spacial cases.

 (1)  While in the asymptotic limit $r\rightarrow\infty$, we have
\begin{eqnarray}
E_{KK} (r\rightarrow\infty)=m\left  (1+\frac{\gamma}{2}\right ),
\end{eqnarray}
which equals the ADM mass of the stationary Kaluza-Klein spacetime
.

 (2)  At the event horizon $r=r_+$, the energy of the stationary
Kaluza-Klein spacetime becomes
\begin{eqnarray}
E_{KK} (r_+)&=&\frac{\sqrt{1+\gamma} (r^2_++a^2)}
{2r_+^{1/2}A^{\frac{1}{4}}}F_1\left  (\frac{1}{2},\frac{1}{4},
\frac{1}{4},\frac{3}{2},-\frac{a^2}{r_+^2},-\frac{a^2}{A}\right)\nonumber\\
&+&\frac{a^2}{24r_+^2}\frac{\sqrt{r_+} (2r_+^2+\gamma r_+^2+\gamma
a^2)}{A^{\frac{3}{4}}}F_1\left
 (\frac{3}{2},\frac{3}{4},\frac{3}{4},\frac{5}{2},-\frac{a^2}{r_+^2},
-\frac{a^2}{A}\right )\nonumber\\
&+&\frac{1}{4}\left\{\sqrt{r_+^2+a^2} (1+\gamma)^\frac{1}{4}
+\frac{\sqrt{r_+} A^{\frac{3}{4}}}{\sqrt{ (1+\gamma)
(r_+^2+a^2)}}\right\} {_2F_1}\left
 (\frac{1}{2},\frac{3}{4},\frac{3}{2},\frac{-\gamma
a^2}{ (1+\gamma)r_+^2}\right ), \label{KKE}
\end{eqnarray}
where $A= (r_+^2+a^2) (1+\gamma)-a^2$ and the Appell
hypergeometric function $F_1 (a,b_1,b_2,c,x,y)$ and the
hypergeometric function $_2F_1 (a,b,c,z)$ are defined by
\begin{eqnarray}
&&F_1 (a,b_1,b_2,c,x,y)=\Sigma^\infty_{m=0}\Sigma^\infty_{m=0}
\frac{ (a)_{m+n} (b_1)_m (b_2)_n}{ (c)_{m+n}}
\frac{x^my^n}{m!n!},\nonumber\\
&&_2F_1 (a,b,c,z)=\Sigma^\infty_{k=0}\frac{ (a)_k (b)_k}{ (c)_k}
\frac{z^k}{k!}.\nonumber
\end{eqnarray}
As $\gamma\rightarrow 0$, Eq.  (\ref{KKE}) shows that $E_{KK}
(r_+)\rightarrow \frac{a\sqrt{r_+^2+a^2}+
(2a^2+3r^2)ArcSinh[\frac{a}{r_+}]}{4a}$, which is just the energy
of the Kerr spacetime at the horizon.

 (3)  If we consider the slow rotation approximation, we get
\begin{eqnarray}
 E_1&=&\frac{1}{4}\int_0^\pi{\sqrt{\rho}\sqrt{\sigma}\sin{\theta}}d\theta \nonumber\\
   &=&\left[\frac{r (1-v^2)}{2mv^2+r (1-v^2)}\right]^\frac{3}{4}
   \frac{6r^2[2mv^2+r (1-v^2)]+a^2 [m (2+v^2)+2r (1-v^2)]}{6r^2 (1-v^2)}
   +
   O\left (\frac{a^4}{r^4}\right),\nonumber\\
 E_2&=&-\frac{1}{4}\int_0^\pi{\frac{\sqrt{\Delta}\ \partial_r{\Sigma}}{\sqrt{\rho}\sqrt{\sigma}}
 \sin{\theta}}d\theta \nonumber\\
   &=&-\frac{\sqrt{r^2-2mr+a^2}}{12r^4 (1-v^2)^2}\left[\frac{r (1-v^2)}{2mv^2+r (1-v^2)}\right]^\frac{7}{4}
   \bigg\{6r^2\left[2mv^2+r (1-v^2)\right ]\left[3mv^2+2r (1-v^2)\right ]\nonumber\\
   &&-a^2\left[m^2v^2 (4+5v^2)+mr (4+7v^2) (1-v^2)+2r^2 (1-v^2)^2\right ]\bigg\}
 +O\left (\frac{a^4}{r^4}\right). \nonumber
\end{eqnarray}
Then, the energy of the slow rotating Kaluza-Klein spacetime is
shown by
\begin{eqnarray}
E_{KK}&=&E_1+E_2\nonumber\\
 &=&\left[\frac{r (1-v^2)}{2mv^2+r (1-v^2)}\right]^\frac{3}{4}
   \left\{\frac{6r^2[2mv^2+r (1-v^2)]+a^2 [m (2+v^2)+2r (1-v^2)]}{6r^2 (1-v^2)} \right \}\nonumber\\
   &&-\frac{\sqrt{r^2-2mr+a^2}}{12r^4 (1-v^2)^2}\left[\frac{r (1-v^2)}{2mv^2+r (1-v^2)}\right]^\frac{7}{4}
   \bigg\{6r^2\left[2mv^2+r (1-v^2)\right ]\left[3mv^2+2r (1-v^2)\right ]\nonumber\\
   &&-a^2\left[m^2v^2 (4+5v^2)+mr (4+7v^2) (1-v^2)+2r^2 (1-v^2)^2\right ]\bigg\}+O\left (\frac{a^4}{r^4}\right),\label{EKKNN}
\end{eqnarray}
which is just the Brown-York quasilocal energy of the Kaluza-Klein
spacetime  \cite{JingE}. It is interesting to note that the term
$-E_1$ corresponds exactly to $Eq. (2.13)$ in Ref. \cite{JingE}
which is a reference term proposed in Ref. \cite{Hawking} for the
normalization of the energy with respect to a reference spacetime,
and the term $E_2$ is Eq. (2.7) in Ref. \cite{JingE}.

\section{ The energy of rotating Cveti\v{c}-Youm spacetime }
The metric of the non-extreme dyonic rotating spacetime  in terms
of the four-dimensional bosonic fields is \cite{JingE} \cite{MCDY}
\begin{eqnarray}
\label{1cy}ds^{2}&=&-\frac{\Delta-a^2\sin^{2}{\theta}}{\sqrt{X}}dt^{2}
-\frac{4ma}{\sqrt{X}}[ (\cosh{\delta_{p1}}\cosh{\delta_{p2}}\cosh{\delta_{e1}}\cosh{\delta_{e2}}\nonumber\\
&-&\sinh{\delta_{p1}}\sinh{\delta_{p2}}\sinh{\delta_{e1}}\sinh{\delta_{e2}})r
+2m\sinh{\delta_{p1}}\sinh{\delta_{p2}}\sinh{\delta_{e1}}\sinh{\delta_{e2}}]\sin^{2}{\theta}dtd\phi\nonumber\\
&+&\frac{\sqrt{X}}{\Delta}dr^{2}+\sqrt{X}d\theta^{2}
+\frac{Y^2}{\sqrt{X}}\sin^{2}{\theta}d\phi^{2},
\end{eqnarray}
with
\begin{eqnarray}
\Delta&=&r^2-2mr+a^2,\nonumber\\
f (r)&=& (r+2m\sinh^2{\delta_{p1}}) (r+2m\sinh^2{\delta_{p2}}) (r+2m\sinh^2{\delta_{e1}}) (r+2m\sinh^2{\delta_{e2}}),\nonumber\\
q_1&=&\sinh^2{\delta_{p1}}+\sinh^2{\delta_{p2}}+\sinh^2{\delta_{e1}}+\sinh^2{\delta_{e2}},\nonumber\\
q_2&=&2\cosh{\delta_{p1}}\cosh{\delta_{p2}}\cosh{\delta_{e1}}\cosh{\delta_{e2}}
\sinh{\delta_{p1}}\sinh{\delta_{p2}}\sinh{\delta_{e1}}\sinh{\delta_{e2}}\nonumber\\
&-&2\sinh^2{\delta_{p1}}\sinh^2{\delta_{p2}}\sinh^2{\delta_{e1}}\sinh^2{\delta_{e2}}
-\sinh^2{\delta_{p2}}\sinh^2{\delta_{e1}}\sinh^2{\delta_{e2}}\nonumber\\
&-&\sinh^2{\delta_{p1}}\sinh^2{\delta_{e1}}\sinh^2{\delta_{e2}}
-\sinh^2{\delta_{p1}}\sinh^2{\delta_{p2}}\sinh^2{\delta_{e2}}-\sinh^2{\delta_{p1}}\sinh^2{\delta_{p2}}\sinh^2{\delta_{e1}},\nonumber\\
W&=&2q_1mr+4q_2m^2+a^2\cos^{2}{\theta},\nonumber\\
X&=&f (r)+ (2a^2r^2+Wa^2)\cos^{2}{\theta}\nonumber\\
&=&f (r)+ (2r^2+2q_1mr+4q_2m^2)a^2\cos^{2}{\theta}+ (a^2\cos^{2}{\theta})^2,\nonumber\\
Y^2&=&f (r)+Wa^2+a^2 (1+\cos^{2}{\theta})r^2+2ma^2r\sin^{2}{\theta}\nonumber\\
&=&f (r)+ (r^2+2mr+2q_1mr+4q_2m^2)a^2+\Delta
a^2\cos^{2}{\theta},\nonumber
\end{eqnarray}
where $\delta_{p1}$, $\delta_{p2}$, $\delta_{e1}$, and
$\delta_{e2}$ are four boosts, and $a$ and $m$ represent the
rotational and mass parameters, respectively.

The $ADM$ mass and angular momentum $J$ of the solution can be
expressed as
\begin{eqnarray}
M&=&\frac{m}{2} (\cosh^2{\delta_{p1}}+\cosh^2{\delta_{p2}}+\cosh^2{\delta_{e1}}+\cosh^2{\delta_{e2}})-m,\nonumber\\
J&=&ma
(\cosh{\delta_{p1}}\cosh{\delta_{p2}}\cosh{\delta_{e1}}\cosh{\delta_{e2}}
-\sinh{\delta_{p1}}\sinh{\delta_{p2}}\sinh{\delta_{e1}}\sinh{\delta_{e2}}).
\end{eqnarray}
The spacetime   has the inner $r_-$ and outer $r_+$ horizons at
\begin{eqnarray}
r_\pm=m\pm\sqrt{m^2-a^2},
\end{eqnarray}
provided $m\geq a$. By using the Eq.  (\ref{def}),  the energy of
the rotating Cveti\v{c}-Youm spacetime  is given by
\begin{eqnarray}\label{ecy}
E_{CY}=\frac{1}{4}\int_0^\pi{\left (X^{\frac{1}{4}}+
\frac{Y}{X^{\frac{1}{4}}}
 -\frac{\sqrt{\Delta}\
 \partial_r{Y}}{X^{\frac{1}{4}}}\right )\sin{\theta}}d\theta.
\end{eqnarray}
The integral  (\ref{ecy}) cannot be expressed in simple functions.
However, we can get the integral for some spacial cases:

 (1)  Taking the asymptotic limit $r\rightarrow\infty$, we obtain
\begin{eqnarray}
E_{CY} (r\rightarrow\infty)=\frac{1}{2} (\cosh^2{\delta_{p1}}
+\cosh^2{\delta_{p2}}+\cosh^2{\delta_{e1}}+\cosh^2{\delta_{e2}})-m,
\end{eqnarray}
which is just the ADM mass of the rotating Cveti\v{c}-Youm
spacetime.

 (2)  At the event horizon $r=r_+$, the energy of the rotating
Cveti\v{c}-Youm black hole  becomes
\begin{eqnarray}
E_{CY} (r_+)&=&\frac{\sqrt{V+\Gamma}}
{2\Gamma^{\frac{1}{4}}}F_1\left  (\frac{1}{2},\frac{1}{4},
\frac{1}{4},\frac{3}{2},-\frac{2a^2}{U+\sqrt{U^2-4\Gamma}},-\frac{2a^2}{U-\sqrt{U^2-4\Gamma}}\right )\nonumber\\
&+&\frac{a^2U}{24\Gamma^{\frac{3}{4}}} F_1\left
 (\frac{3}{2},\frac{3}{4},\frac{3}{4},\frac{5}{2},
-\frac{2a^2}{U+\sqrt{U^2-4\Gamma}},-\frac{2a^2}{U-\sqrt{U^2-4\Gamma}}\right )\nonumber\\
&+&\frac{1}{8}\left [2 (a^4+a^2U+\Gamma)^{\frac{1}{4}}
+\frac{\sqrt{2}\Gamma^{\frac{1}{4}}\sqrt{a^2U-a^2\sqrt{U^2-4\Gamma}+2\Gamma}}{\sqrt{a^4+a^2U+\Gamma}}\right ]\nonumber\\
&&_2F_1\left
 (\frac{1}{2},\frac{3}{4},\frac{3}{2},\frac{2a^2\sqrt{U^2-4\Gamma}}{a^2U+a^2
 \sqrt{U^2-4\Gamma}+2\Gamma}\right),
\end{eqnarray}
where $ \Gamma=f (r_+),$ $ U=2r_+^2+2q_1mr_++4q_2m^2 $  and $ V=
(r_+^2+2mr+2q_1mr_++4q_2m^2)a^2.$

While $\delta_{p1}=\delta_{p2}=\delta_{e1}=\delta_{e2}= 0$,
$E_{CY} (r_+)=\frac{a\sqrt{r_+^2+a^2}+
(2a^2+3r^2)ArcSinh[\frac{a}{r_+}]}{4a}$, which is just the energy
of the Kerr black hole at the horizon.

 (3)  In the slowly rotating approximation, we find the energy of the
Cveti\v{c}-Youm spacetime is $E=E_1+E_2$, in which
\begin{eqnarray}\label{e1cy}
E_1&=&\frac{1}{4}\int_0^\pi  ( {X^{\frac{1}{4}}+
\frac{Y}{X^{\frac{1}{4}}} )
 \sin{\theta}}d\theta \nonumber\\
 &=&\frac{1}{6f (r)}[6f (r)+a^2 (2r^2+2mr+3q_1mr+6q_2m^2)]+O (\frac{a^4}{r^4}),\\
 E_2&=&-\frac{1}{4}\int_0^\pi{\frac{\sqrt{\Delta}
\partial_r{Y}}{X^{\frac{1}{4}}}\sin{\theta}}d\theta\nonumber\\
&=&-\frac{1}{8}\int_0^\pi{\frac{\sqrt{r^2-2mr+a^2}[\partial_r{f
(r)+2 (r+m+q_1m)+2 (r-m)a^2\cos^2{\theta}}]} {X^{\frac{1}{4}}\sqrt{f
(r)+a^2 (r^2+2mr+2q_1mr+4q_2m^2)+
(r^2-2mr+a^2)a^2\cos^2{\theta}}}\sin{\theta}}d\theta. \label{e2cy}
\end{eqnarray}
It is interesting to note that Eq. (\ref{e1cy}) corresponds to Eq.
(3.13) in Ref. \cite{JingE} and Eq. (\ref{e2cy}) corresponds to
Eq. (3.8) in Ref. \cite{JingE}.

\section{Conclusion}

In this paper, the field equation with the cosmological constant
is derived and the energy of the general 4-dimensional stationary
axisymmetric spacetime is studied in the context of the
hamiltonian formulation of the teleparallel equivalent of general
relativity  (TEGR). The main result is that a general energy
expression for the asymptotically flat/de Sitter/Anti-de Sitter
stationary spacetimes in the Boyer-Lindquist coordinate is
obtained by means of the integral form of the constraints
equations of the formalism naturally without any restriction on
the metric parameters. It is surprised to learn that the energy
expression for this spacetime is very simple and relevant to the
metric components $g_{rr}$, $g_{\theta\theta}$ and $g_{\phi\phi}$
only. As examples, by using this formula we calculate the energies
of the KN, KN-AdS, Kaluza-Klein, and Cveti\v{c}-Youm spacetimes.
By comparing the results with known results, we find that our
results are exactly agree with  the values obtained by using other
methods for some special cases.

\vspace*{0.20cm}

\begin{acknowledgments}This work was supported by the
National Natural Science Foundation of China under Grant No.
10473004; the FANEDD under Grant No. 200317; and the SRFDP under
Grant No. 20040542003; the Hunan Provincial Natural Science
Foundation of China under Grant No.  05JJ20001.
\end{acknowledgments}

\appendix

\section{  Calculation of torsion and momenta}

We present the nonvanishing components of the torsion tensor
related to the set of tetrads given in Eq. ( \ref{matric-e}) as
following:
\begin{eqnarray}
{\stackrel{*}{T}}{}_{ (0)01} &= &\partial_r\; A, \nonumber\\
{\stackrel{*}{T}}{}_{ (0)02} &= &\partial_\theta\; A, \nonumber\\
{\stackrel{*}{T}}{}_{ (1)01} &= &-\sin{\theta}\sin{\phi}\;\partial_r\;{B}, \nonumber\\
{\stackrel{*}{T}}{}_{ (1)02} &= &-\sin\phi\;\partial_\theta\; ({B\sin\theta}), \nonumber\\
{\stackrel{*}{T}}{}_{ (1)03} &= &-B\sin{\theta}\cos{\phi}, \nonumber\\
{\stackrel{*}{T}}{}_{ (1)12} &= &\cos\theta\cos\phi\;\partial_r\;{
(Dr)}
- cos\phi\;\partial_\theta\;{ (C\sin\theta)}, \nonumber\\
{\stackrel{*}{T}}{}_{ (1)13} &= & \sin\theta\sin\phi [C
-\partial_r{ (Fr)}], \nonumber\\
{\stackrel{*}{T}}{}_{ (1)23} &= &Dr\cos\theta\sin\phi
- \sin\phi\;\partial_\theta\;{ (Fr \sin{\theta})}, \nonumber\\
{\stackrel{*}{T}}{}_{ (2)01} &= & \sin{\theta}\cos\phi\;\partial_r\;{B}, \nonumber\\
{\stackrel{*}{T}}{}_{ (2)02} &= & \cos\phi\;\partial_\theta\;{ (B\sin\theta)}, \nonumber\\
{\stackrel{*}{T}}{}_{ (2)03} &= & -B\sin\theta\;\sin\phi, \nonumber\\
{\stackrel{*}{T}}{}_{ (2)12} &= &
\cos\theta\sin\phi\;\partial_r\;{ (Dr)}
- \sin\phi\;\partial_\theta\;{ (C\sin\theta))}, \nonumber\\
{\stackrel{*}{T}}{}_{ (2)13} &= & -\sin\theta\cos\phi [C
-\partial_r{ (Fr)}], \nonumber\\
{\stackrel{*}{T}}{}_{ (2)23} &= &\cos\phi\;\partial_\theta\;{
(Fr\sin{\theta})}
-Dr\cos\theta\cos\phi, \nonumber\\
{\stackrel{*}{T}}{}_{ (3)12} &= & -\sin\theta\;\partial_r\;{ (Dr)}
-
\partial_\theta\; (C\cos\theta).
\end{eqnarray}
Then, using Eq. ( \ref{expession1}) we find  that  the component
$\Pi^{ (0)1}$ of the canonical momentum for the stationary
axisymmetric  spacetime can be expressed as
\begin{eqnarray}
\Pi^{ (0)1}=-2k[sin\theta
\sqrt{g_{\theta\theta}}+\sqrt{g_{\phi\phi}}-\frac{\partial_r{
(\sqrt{g_{\theta\theta}} \sqrt{g_{\phi\phi}})}}{\sqrt{g_{rr}}}].
\end{eqnarray}

\end{document}